\documentclass{article} % use larger type; default would be 10pt

\usepackage[utf8]{inputenc} % set input encoding (not needed with XeLaTeX)
\usepackage[T1]{fontenc}

\usepackage{geometry} % to change the page dimensions
\usepackage{graphicx} % support the \includegraphics command and options
\usepackage{amsmath}
\usepackage{amssymb}
\usepackage{booktabs} % for much better looking tables
\usepackage{array} % for better arrays (eg matrices) in maths
\usepackage{paralist} % very flexible & customisable lists (eg. enumerate/itemize, etc.)
\usepackage{verbatim} % adds environment for commenting out blocks of text & for better verbatim
\usepackage{subfig} % make it possible to include more than one captioned figure/table in a single float
\usepackage{float}
\usepackage{color}
\usepackage{fancyhdr} % This should be set AFTER setting up the page geometry
\usepackage{cite}

\usepackage{sectsty}
\allsectionsfont{\sffamily\mdseries\upshape} % (See the fntguide.pdf for font help)
\usepackage[nottoc,notlof,notlot]{tocbibind} % Put the bibliography in the ToC
\usepackage[titles,subfigure]{tocloft} % Alter the style of the Table of Contents

\newtheorem{example}{Example}
\newtheorem{remark}{Remark}

\title{Remarks on entanglement and identical particles}

\author{F. Benatti$^{a,b}$, R. Floreanini$^b$, F. Franchini$^c$, U. Marzolino$^c$ \\
\\
\small ${}^a$ Universit\`a di Trieste, I-34151 Trieste, Italia \\
\small ${}^b$ Istituto Nazionale di Fisica Nucleare, Sezione di Trieste, I-34151 Trieste, Italia \\
\small ${}^c$Institut Ru\dj er Bo\v{s}kovi\'c, HR-10000 Zagreb, Hrvatska}

\date{\empty}
\begin{document}

\maketitle

\begin{abstract}
We argue that in the case of identical particles the most natural identification of separability, that is of absence of non-classical correlations, is via the factorization of mean values of commuting observables. It thus follows that separability and entanglement depend both on the state and on the choice of observables and are not absolute notions. We compare this point of view with a recent novel approach to the entanglement of identical particles, which allows for the definition of an entanglement entropy from a suitably defined reduced particle density matrix, without the need of labeling the system constituents. We contrast this figure of merit with the aforementioned lack of an absolute notion of entanglement by considering few paradigmatic examples.
\vskip .5cm

\end{abstract}

\section{Introduction}

For distinguishable particles the notion of entanglement has a well defined and universally accepted formulation~\cite{Horodecki}; however, this is not the case for indistinguishable 
particles, where the debate is still open on whether the required symmetrization or anti-symmetrization of identical particle states can be by itself a source of entanglement as a usable tool for quantum technologies \cite{Zanardi2001, Schliemann2001, Paskauskas2001, Li2001, Ghirardi2002, 
Shi2003, Wiseman2003, Barnum2004, Sasaki2011, Tichy2013}, see also the recent review~\cite{Braun}. Typically, what one tries to do is to export to the case of Bosons and Fermions what one thinks to be a fundamental mark of entanglement in the case of distinguishable particles. 

Taking as a mark of separability for pure states, namely of absence of entanglement, the lack of correlations among commuting observables, one can consistently formulate a theory of entanglement for identical particles systems, and of its applications, using only the tools of the second quantization approach to quantum statistical mechanics~\cite{Zanardi2001,Barnum2004,Benatti2010,Benatti2011,Argentieri2011,Benatti2012,
Benatti2012-2,Marzolino2013,Benatti2014-4,Benatti2014,Benatti2014-2,Benatti2014-3,
Marzolino2015,Marzolino2016,Benatti2016}. The resulting evidence is that separability and entanglement are not absolute notions as they depend on the sets of commuting observables that are chosen. Mathematically, these sets have the structure of an algebra, and more specifically of a $C^*$-algebra \cite{BratteliRobinson,Strocchi}. In other terms, the same state can be seen both as separable with respect to two commuting sub-algebras and as entangled with respect to a different pair of commuting sub-algebras. This is also true for distinguishable particles; however, in that case there is a privileged choice of commuting sub-algebras, namely the ones associated with each specific particle. This attribution of observables to specific particles is not possible anymore in the case of identical particles, so that the need and also the opportunity of referring to far wider classes of commuting sub-algebras and correlations naturally emerge.

Recently, a new approach to the description of identical particle systems has been proposed~\cite{LoFranco2016,LoFranco2017,Bellomo2017}: through a modification of the first quantization approach to identical particle systems, states are constructed without assigning particle labels to single particle states and thus without the need to symmetrize or anti-symmetrize tensor product states. Instead, in this formulation, which we shall refer to as the \textit{no-label approach}, the many-body structure is encoded through an original definition of the scalar products between states (with the same number of particles). Moreover, within such a context, a projection-like operation from higher to lower particle number sectors of the Fock space can be constructed and used to firstly derive a one particle-like reduced density matrix from pure states of two identical particles and, secondly, to define the amount of their entanglement as the von Neumann entropy of the reduced one-particle density matrix, exactly as one does in the case of two distinguishable particles. Particular applications of this approach regard the attribution of entanglement  to bipartite states of two identical particles with a spatial and an internal degrees of freedom by localizing them in space.

In the following we first motivate the approach to identical particle entanglement based on factorization of correlation functions, which we also briefly summarize, then we provide a short overlook of the no-label approach and finally we compare the conclusions of the two approaches in a few simple concrete examples.

\section{Separability and entanglement for distinguishable particles}

In the common lore, quantum entanglement stands for correlations without classical counterparts~\cite{Horodecki}: these non-classical correlations set the ground for technological performances otherwise impossible basing on purely classical means~\cite{Nielsen,Benatti1}. In the case of a classical system, the simplest correlations are those embodied by two-point expectation values: these involve the averages 
\begin{equation}
\label{classcorr}
\mu(X_1X_2)=\int_{\mathcal{X}}{\rm d}\mu(s) \, X_1(s)\,X_2(s)
\end{equation}
of products of two stochastic variables $X_{1,2}(s)$ over a suitable measure space $\mathcal{X}$ with respect to a probability distribution $\mu(s)$.
Through~\eqref{classcorr}, the probability distribution $\mu(s)$ assigns mean values to all possible stochastic variables and therefore completely characterises the statistical properties of the system or, in more physical words, the probability $\mu(s)$ exhaustively defines the \textit{state} of classical systems.
For instance, absence of correlations between two stochastic variables $X_{1,2}(s)$ 
relative to a given state $\mu(s)$ is described by the factorization 
\begin{equation}
\label{classsep}
\mu(X_1\,X_2)=\mu(X_1) \, \mu(X_2)\ .
\end{equation}

In quantum mechanics, the probability distribution $\mu$ is usually replaced by a density 
matrix, or mixed state, $\varrho$ acting on a Hilbert space $\mathbb{H}$, that is by a convex combination 
\begin{equation}
\label{densmat}
\varrho=\sum_j\lambda_j\,\vert\psi_j\rangle\langle\psi_j\vert\ ,\qquad \lambda_j\geq0\ ,\sum_j\lambda_j=1\ .
\end{equation}
of weighted one-dimensional projections $\vert\psi_j\rangle\langle\psi_j\vert$, the so-called pure states: these provide complete descriptions of the quantum systems states. Mixed states instead correspond to statistical ensembles distributed according to the given weights, whereby two-point correlation functions in~\eqref{classcorr} then read
\begin{eqnarray}
\label{qcorr}
{\rm Tr}(\rho\,A\,B)=\sum_j\lambda_j\,\langle\psi_j\vert A\,B|\psi_j\rangle\ ,
\end{eqnarray}
where $A=A^\dag$ and $B=B^\dag$ are observables that is self-adjoint operators on $\mathbb{H}$. If $A$ and $B$ commute, $AB=BA$ is an observable in analogy with the integrand of \eqref{classcorr}; moreover, the statistics of the outcomes of measurements of $A$ and $B$ are independent of each other. The statistical independence ensuing from commutativity must then correspond to absence 
of correlations among the observables $A$ and $B$ with respect to a pure state $\vert\psi\rangle\langle \psi\vert$. This is equivalent to the request that the corresponding two-point correlation function factorize
\begin{equation}
\label{qsep}
\langle\psi\vert A\,B|\psi\rangle=\langle\psi\vert A|\psi\rangle\langle \, \psi\vert B|\psi\rangle\ .
\end{equation} 
Of course, for a generic mixed state $\varrho$, ${\rm Tr}(\rho\,A\,B)\neq{\rm Tr}(\rho\,A)\,{\rm Tr}(\rho\,B)$, even in absence of quantum correlations in the constituent pure states; however, this lack of factorization is due to the classical correlations carried by the weights $\lambda_j$ in~\eqref{densmat}. 
 
As far as separability and entanglement of distinguishable particles are concerned, these are concepts related to a natural notion of locality embodied by (local) quantum observables of bipartite quantum systems $S_1+S_2$ that are typically tensor products $O_1\otimes O_2$ of observables $O_1$ pertaining to system $S_1$, that is acting on the Hilbert space $\mathbb{H}_1$ of $S_1$, and $O_2$ pertaining to system $S_2$, that is acting on the Hilbert space $\mathbb{H}_2$ of $S_2$. The observables $O_{1,2}$ are embedded within the algebra of operators of $S_1+S_2$ as $\tilde O_1=O_1\otimes 1$ and $\tilde O_2=1\otimes O_2$ so that $\tilde O_1\tilde O_2=O_1\otimes O_2$. Then, pure states $\vert\Phi_{12}\rangle$ carrying no correlations with respect to $O_{1,2}$ are such that
\begin{equation}
\label{1}
\langle\Phi_{12}\vert\tilde O_1\tilde O_2\vert\Phi_{12}\rangle=\langle\Phi_{12}\vert\tilde O_1\vert\Phi_{12}\rangle\, \langle\Phi_{12}\vert\tilde O_2\vert\Phi_{12}\rangle\ .
\end{equation}
These states are known as \textit{separable vector states} and are of the form $\vert\Phi_{12}\rangle=\vert\phi_1\rangle\otimes\vert\phi_2\rangle$. Indeed, 
if $\vert\Phi_{12}\rangle=\vert\phi_1\rangle\otimes\vert\phi_2\rangle$, the above factorization occurs for all possible choices of observables $O_{1,2}$. 
Vice versa, consider the Schmidt decomposition 
$$
\vert\Phi_{12}\rangle=\sum_j\lambda_j\vert\phi^{(1)}_j\rangle\otimes\vert\phi^{(2)}_j\rangle\ ,
$$
with Schmidt coefficients $\lambda_j>0$ and orthogonal vectors 
$\langle\phi_k^{(a)}\vert\phi^{(a)}_\ell\rangle=\delta_{k\ell}$, $a=1,2$.
By choosing in~\eqref{1} $O_1=\vert\phi^{(1)}_i\rangle\langle\phi^{(1)}_i\vert$ and 
$O_2=\vert\phi^{(2)}_j\rangle\langle\phi^{(2)}_j\vert$, it follows that
$\lambda_i\lambda_j=\lambda_i^2\lambda_j^2$ for all possible pairs of Schmidt coefficients. Together with the normalization condition $\sum_j\lambda_j^2=1$, this implies that only one of them can be different from zero. Therefore, if $\vert\Phi_{12}\rangle$ satisfies~\eqref{1} for all possible single particle observables $O_{1,2}$, it must be of  the form $\vert\Phi_{12}\rangle=\vert\phi_1\rangle\otimes\vert\phi_2\rangle$.

Pure states for which factorization fails for at least one pair of observables $O_{1,2}$ as defined above  are then \textit{entangled}. 
In the case of mixed quantum states, being separable, that is non-entangled, means being a convex combination of separable states
\begin{equation}
\label{3}
\rho_{\textnormal{sep}}=\sum_j\lambda_j\,\vert\psi_j^{(1)}\rangle\langle\psi_j^{(1)}\vert\otimes\vert\psi_j^{(2)}\rangle\langle\psi_j^{(2)}\vert\ ,\quad\sum_j\lambda_j=1\ ,\ \lambda_j\geq 0\ ,
\end{equation}
so that
\begin{equation}
\label{2b}
{\rm Tr}\Big(\rho_{\textnormal{sep}}\, O_1\otimes O_2\Big)=\sum_j\lambda_j\,\langle\psi^{(1)}_j\vert\,O_1\vert\psi^{(1)}_j\rangle\,\langle\psi^{(2)}_j\vert\,O_2\,\vert\psi^{(2)}_j\rangle\ ,\quad \forall\ O_{1,2}\ ,
\end{equation}
whereby correlations are only those supported by the weights $\lambda_j$.

\begin{remark}
\label{rem1}
The factorized form of separable pure states is clearly associated with the tensor product structure, $\mathbb{H}=\mathbb{H}_1\otimes\mathbb{H}_2$ of the Hilbert space $\mathbb{H}$ of the compound system $S_1+S_2$.  
In this standard context, the locality of observables of the form $O_1\otimes O_2$ is implicitly understood with respect to the given tensor product structure of the Hilbert space $\mathbb{H}$. Their locality is a particular instance of algebraic independence, namely of commutativity: it is also of particle type, since $O_{1,2}$ are observables that can be attributed to degrees of freedom identifiable with different particles. We will refer to this kind of locality as \textit{particle locality}.
\end{remark}

The following example shows that, by allowing for more general algebraic independence than particle-locality, even in the standard settings, is not possible to establish the absence of correlations independently from the context, {\it i.e. from the choice of the sub-algebras}. For sake of simplicity, let $S_{1,2}$ be finite level systems, so that $\mathbb{H}_1=\mathbb{C}^{d_1}$ and $\mathbb{H}_2=\mathbb{C}^{d_2}$ and local observables $O_1\otimes O_2$ are constructed from $O_1=O^\dag_1$ belonging to the sub-algebra $M_{d_1}(\mathbb{C})$ of $d_1\times d_1$ matrices, respectively $O_2=O_2^\dag$ from the algebra $M_{d_2}(\mathbb{C})$ of $d_2\times d_2$ matrices.

\begin{example}
\label{ex1}
Let $d_1=d_2=2$: the statistical independence of the local observables $O_{1,2}$ fails in general for the (maximally entangled) Bell states:
\begin{eqnarray}
\label{4a}
\vert\Psi_{\pm}\rangle&=&\frac{1}{\sqrt{2}}\Big(\vert0\rangle\otimes\vert1\rangle\pm\vert1\rangle\otimes\vert 0\rangle\Big)\\
\label{4b}
\vert\Phi_{\pm}\rangle&=&\frac{1}{\sqrt{2}}\Big(\vert0\rangle\otimes\vert0\rangle\pm
\vert1\rangle \otimes\vert 1\rangle\Big)\ ,
\end{eqnarray}
with $\vert 0\rangle$ and $\vert 1\rangle$ the orthonormal basis in $\mathbb{C}^2$ given by the eigenstates of $\sigma_3$: $\sigma_3\vert k\rangle=(-)^k\vert k\rangle$, $k=0,1$.
Indeed, choosing $O_{1,2}=\sigma_3$, one gets
\begin{equation}
\label{5}
\langle\Psi_+\vert\sigma_3\otimes\sigma_3\vert\Psi_+\rangle=-1\neq
\langle\Psi_+\vert\sigma_3\otimes1\vert\Psi_+\rangle\,\langle\Psi_+\vert1\otimes\sigma_3\vert\Psi_+\rangle=0\ .
\end{equation}
Let now $\mathcal{A}_+$, respectively $\mathcal{A}_-$ be the commutative sub-algebras of $M_2(\mathbb{C})\otimes M_2(\mathbb{C})$ generated by the identity and the projectors $P^\Psi_+$ and $P^\Phi_+$ onto $\vert\Psi_+\rangle$ and $\vert\Phi_+\rangle$, respectively by the identity and the projectors $P^\Psi_-$ and $P^\Phi_-$ onto $\vert\Psi_-\rangle$ and $\vert\Phi_-\rangle$. The two sub-algebras $\mathcal{A}_\pm$ commute and their expectation values with  respect to (\ref{4a},\ref{4b}) factorize, indicating that from this point of view even the Bell states can be identified as {\it not entangled}. At the same time, the expectations with respect to $\vert\Psi\rangle=\vert0\rangle\otimes\vert0\rangle$, which factorize with respect to the \textit{particle-local} pair $\Big(M_2(\mathbb{C})\otimes 1,1\otimes M_2(\mathbb{C})\Big)$, do not factorize with respect to $(\mathcal{A}_+,\mathcal{A}_-)$. Indeed,
\begin{equation}
\label{6}
\langle\Psi\vert P^\Phi_+\,P^\Phi_-\vert\Psi\rangle=0\neq
\langle\Psi\vert P^{\Phi}_+\vert\Psi\rangle\,\langle\Psi\vert P^\Phi_-\vert\Psi\rangle=\frac{1}{4}\ .
\end{equation}
Therefore, $\vert\Psi\rangle$ is separable with respect to particle-locality, while it is no longer separable with respect to the algebraically independent sub-algebras $\mathcal{A}_\pm$.
Clearly, the observables
$$
P^\Phi_\pm=\frac{1}{2}\Big(\vert 0\rangle\langle0\vert\otimes\vert0\rangle\langle0\vert+\vert 1\rangle\langle1\vert\otimes\vert1\rangle\langle1\vert\pm\vert 0\rangle\langle0\vert\otimes\vert1\rangle\langle1\vert\pm\vert 1\rangle\langle1\vert\otimes\vert0\rangle\langle0\vert\Big)
$$
are sums of tensor products and thus particle non-local.
\end{example}

From these considerations, one sees, even in the simplest possible setting, that,  
by abandoning the notion of locality associated with particle identification, and referring instead to algebraic independence, namely to commuting pairs of sub-algebras, then separability and entanglement immediately become notions that need to be specified in relation to the specific context provided by the chosen pairs of commuting sub-algebras.
Obviously, in the standard, distinguishable particle case, one is mainly concerned  with particle-locality because particles can be naturally identified.
However, this is not true anymore for identical particles and one is naturally forced to replace particle-locality by algebraic-independence.

\section{Identical particle entanglement}

In this section, we consider the case of identical, namely indistinguishable particles.
Clearly, the fact that their Hilbert spaces must be either symmetrized (Bosons) or anti-symmetrized (Fermions) hampers the emergence of a unique (natural) tensor product structure for the Hilbert space over which to base the notion of separability and entanglement. 
Of the many different possible approaches to the problem that appeared in the literature in the last years, we will consider and compare two of them. 
The first one~\cite{Benatti2010,Benatti2014,Benatti2012,Benatti2012-2} bases the notion of separability on the absence of classical correlations with respect to products of pairs of commuting observables, while the second one~\cite{LoFranco2016,LoFranco2017,Bellomo2017} makes use of a suitably constructed one particle-reduction of identical particle states. 

\subsection{Second quantization approach}
\label{sec3.1}

Identical particles are most suitably addressed by means of second quantization which is based on the introduction of a single-particle Hilbert space $\mathbb{H}$ and the corresponding creation, $a^\dagger(f)$, and annihilation, $a(g)$, operators, with single-particle vector states $f,g\in\mathbb{H}$, the so-called \textit{modes}. They satisfy Canonical Commutation or Anticommutation Relations:
\begin{equation}
\label{7}
\Big[a(f)\,,\,a^\dag(g)\Big]=\langle f\vert g\rangle\quad\hbox{(CCR)}\quad,\qquad
\Big\{a(f)\,,\,a^\dag(g)\Big\}=\langle f\vert g\rangle\quad\hbox{(CAR)}\ ,
\end{equation}
while all other commutators and anti-commutators vanish.
Polynomials in these operators construct the algebra $\mathcal{A}$ of operators that are used to describe many-body systems consisting of Bosons and Fermions.
Fermionic and Bosonic states amount to positive, normalized linear functionals
\begin{equation}
\label{8}
\omega:\mathcal{A}\mapsto \mathbb{C}\ ,\quad \omega(x^\dag x)\geq 0\quad\forall x\in\mathcal{A}\ ,\quad \omega(1)=1\ .
\end{equation}
We refer to $\omega$ as a {\it state} in that it generalizes the standard (pure and mixed)quantum states given by 
$O\mapsto {\rm Tr}(\rho\,O)$ where $\rho$ is any density matrix.
Pure states are those expectation functionals that cannot be convexly decomposed into weighted sums of other expectation functionals over $\mathcal{A}$: the simplest instance of pure state on either Bosonic or Fermionic $\mathcal{A}$ is provided by the vacuum vector
$\vert0\rangle$ such that 
$$
\omega(a(f))=\langle0\vert\, a(f)\vert0\rangle=0,
\quad
\omega(a^\dag(f))=\langle0\vert\, a^\dag(f)\vert0\rangle=0,
\quad\forall\,f\ .
$$
Two-point correlation functions of the form $\omega(x_1x_2)$ give access to the spatial correlations among observables $x_{1,2}$ localized within volumes $V_{1,2}$.
These observables are in general self-adjoint polynomials in creation and annihilation operators $a(f)$, $a^\dag(f)$ with $f$ supported within either volume $V_1$ or $V_2$.
In the Bosonic case, these polynomials generate sub-algebras $\mathcal{A}_{1,2}$ that, because of~\eqref{7}, commute when $V_1$ and $V_2$ do not overlap. 
When $\omega$ is pure, that is when it cannot be convexly decomposed, then factorization
\begin{equation}
\label{9}
\omega(x_1x_2)=\omega(x_1)\, \omega(x_2)\qquad \forall\ x_1\in\mathcal{A}_1\ ,\ x_2\in\mathcal{A}_2\ ,
\end{equation}
shows absence of spatial correlations; namely, that $\omega$ is spatially-separable, while 
\begin{equation}
\label{9bis}
\omega(x_1x_2)\neq\omega(x_1)\, \omega(x_2)
\end{equation}
for some $x_1\in\mathcal{A}_1\ ,\ x_2\in\mathcal{A}_2$, reveals non-local spatial correlations, or, equivalently, that $\omega$ is spatially entangled (entangled with respect to spatial modes).
By changing modes, for instance by considering single particle vector states that are superpositions of those localized within $V_1$ and those localized within $V_2$,
then one can construct a pair of commuting subalgebras $(\mathcal{B}_1,\mathcal{B}_2)$
different from $(\mathcal{A}_1,\mathcal{A}_2)$, with respect to which a spatially-separable state may easily become entangled.

In Fermionic systems, elements of different sub-algebras can also anticommute in accordance with microcausality \cite{Benatti2014,Benatti2014-3,Benatti2016}.

\begin{example}
\label{ex2}
The simplest physical instance where the previous arguments could be put to test 
is provided by $N$ condensed (Bosonic) cold atoms confined in a two-well potential.
The single particle Hilbert space is $\mathbb{C}^2$ spanned by two vectors, $\{\vert L\rangle,\vert R\rangle\}$, representing an atom being localized within the left, respectively right well. 
Spatial-locality is associated with the independent pair of commuting sub-algebras given by the two sub-algebras $\mathcal{A}_{L,R}$ generated by the creation and annihilation operators $a_L,a^\dag_L$ and $a_R,a^\dag_R$ such that
$$
a^\dag_L\vert0\rangle=\vert L\rangle\ ,\quad a^\dag_R\vert0\rangle=\vert R\rangle\ .
$$
The two sub-algebras commute because of the CCR~\eqref{7}; they have in common only the identity operator and, together, they generate the entire Bosonic algebra $\mathcal{A}$ over the single particle Hilbert space $\mathbb{C}^2$. In the $N$-particle sector, spatially-separable pure states are the Fock number states with $k$ particles in the left well and $N-k$ particles in the right well:
\begin{equation}
\vert k\rangle:=\frac{(a_L^\dag)^k}{\sqrt{k!}}\,\frac{(a^\dag_R)^{N-k}}{\sqrt{(N-k)!}}\,\vert 0\rangle\ .
\label{kdef}
\end{equation}
One can check that, if $P(a_L,a_L^\dag)$ and $Q(a_R,a^\dag_R)$ are polynomials in the left and right creation and annihilation operators, then 
$$
\langle k\vert\,P(a_L,a^\dag_L)\,Q(a_R,a^\dag_R)\,\vert k\rangle= \langle k\vert\,P(a_L,a^\dag_L)\,\vert k\rangle\, \langle k\vert\,Q(a_R,a^\dag_R)\,\vert k\rangle
$$
Indeed, by using the canonical commutation relations, one can rewrite 
\begin{eqnarray*}
\frac{1}{(N-k)!}\,a_R^{N-k}\,Q(a_R,a^\dag_R)\,(a_R^\dag)^{N-k}&=& 
Q_0+Q_1\ ,\\
\frac{1}{k!}a_L^k\,P(a_L,a^\dag_L)\,(a_L^\dag)^k=&=& P_0+P_1\ ,
\end{eqnarray*}
where $Q_0$ and $P_0$ are sums of products $a_{R,L}^n(a_{R,L}^\dag)^n$ of same powers $n\geq 0$ of annihilation and creation operators, while $Q_1$ and $P_1$ contain terms with different powers.  Since $[Q_a,P_b]=0$, $a,b=0,1$, and $Q_1\vert 0\rangle=P_1\vert0\rangle=0$ , one gets
\begin{eqnarray*}
\langle k\vert\,P(a_L,a^\dag_L)\,Q(a_R,a^\dag_R)\,\vert k\rangle
&=&\langle 0\vert Q_0\,P_0\vert 0\rangle=\langle 0\vert\,Q_0\,\vert 0\rangle\,\langle 0\vert\,P_0\,\vert 0\rangle\\
&=&\langle k\vert\,P(a_L,a^\dag_L)\,\vert k\rangle\,\langle k\vert\,Q(a_R,a^\dag_R)\,\vert k\rangle\ .
\end{eqnarray*}
The previous form of pure states is not only sufficient for being spatially-separable, but also necessary. Indeed, one can show that any pure separable state on 
$\mathcal{A}$ is of the form $\mathcal{P}(a^\dag_L)\mathcal{Q}(a^\dag_R)\vert 0\rangle$ with suitable polynomials $\mathcal{P}$ and $\mathcal{Q}$ in the left, respectively right creation operators.

The Bogoliubov transformation
$\displaystyle
b_\pm=\frac{a_L\pm a_R}{\sqrt{2}}$
provides creation and annihilation operators associated with the delocalized modes $\displaystyle\vert E_\pm\rangle=\frac{\vert L\rangle\pm\vert R\rangle}{\sqrt{2}}$. 
The latter states approximate the eigenstates of the single particle double-well Hamiltonian
$$
H=E\,\Big(\vert L\rangle\langle L\vert-\vert R\rangle\langle R\vert\Big)\,+\,U\,\Big(\vert L\rangle\langle R\vert\,+\,\vert R\rangle\langle L\vert\Big)\ ,
$$ 
with $0\leq E\ll U$.  
The new modes $b_\pm$ thus provide an energy-local bipartition consisting of the two commuting sub-algebras
$\displaystyle\mathcal{B}_\pm=\{b_\pm,b_\pm^\dag\}$ with respect to which the spatially-separable state $\vert k\rangle$ in~\eqref{kdef}) is no longer separable as it is a sum of states of the form 
$\displaystyle \frac{(b_-^\dag)^k}{\sqrt{k!}}\,\frac{(b_+^\dag)^{N-k}}{\sqrt{(N-k)!}}\,\vert 0\rangle$. 
\end{example}

\subsection{Particle entanglement without particle labels}

In a recent approach to identical particles entanglement aiming at avoiding any statement based on the erroneous labeling of particles that cannot be identified, a 
one-particle reduction operation from the state of two identical particles to one was introduced, thereby allowing for the quantification of identical particle entanglement 
by means of the von Neumann entropy of the so obtained reduced density matrix. We briefly review this no-label approach.

This proposal is neither a first quantization approach making use of (anti)-symmetrized
Hilbert space vectors, nor a second quantization approach based on mode creation and annihilation operators. Instead, it is derived from the standard Hilbert space structure 
of single particles and from the following construction of two-particle states and their scalar products~\footnote{Though extendible to more than two particles, we shall however restrict our considerations to the simplest possible non-trivial scenario.}.
Given single particle vector states $\vert\phi_{1,2}\rangle$ in a Hilbert space $\mathbb{H}$, two-particle vector states are generated from linear combinations of all pairs $(\phi_1,\phi_2)$, denoted as $\vert \phi_1,\phi_2\rangle$ among which a scalar product is defined by  
\begin{equation}
\label{LFscprod0}
\langle\phi_1,\phi_2\vert\phi_1',\phi_2'\rangle_\eta=\langle\phi_1\vert\phi_1'\rangle\,\langle\phi_2\vert\phi_2'\rangle\,+\,\eta\,\langle\phi_1\vert\phi_2'\rangle\,\langle\phi_2\vert\phi_1'\rangle\ ,
\end{equation}
where $\langle\phi\vert\phi'\rangle$ is the usual single particle scalar product, while $\eta=1$ for Bosons and $\eta=-1$ for Fermions. Though the result is the same as with (anti-)symmetrized tensor products of single particle vector states, in this case there is no need for labelling the states with particle indices and then proceed to symmetrization or antisymmetrization.
From direct application of~\eqref{LFscprod0} there follows linearity, \textit{e.g.} $\alpha\vert\phi_1,\phi_2\rangle=\vert\alpha\phi_1,\phi_2\rangle=\vert\phi_1,\alpha\phi_2\rangle$, for all $\alpha\in\mathbb{C}$, and that
\begin{equation}
\label{parity}
\langle\phi_1,\phi_2\vert\phi_1',\phi_2'\rangle_\eta=\eta\,\langle\phi_1,\phi_2\vert\phi_2',\phi_1'\rangle_\eta
\end{equation}
for all $\phi_1,\phi_2\in\mathbb{H}$, so that, under vector exchange, 
$\vert\phi_1,\phi_2\rangle=\eta\,\vert\phi_2,\phi_1\rangle$.
The states $|\phi_1,\phi_2\rangle$ correspond, in the usual first quantization formalism endowed with tensor product between individual particle Hilbert spaces, to the following unnormalized states
\begin{equation}
\frac{1}{\sqrt{2}}\big(|\phi_1\rangle\otimes|\phi_2\rangle+\eta|\phi_2\rangle\otimes|\phi_1\rangle\big)\ ,
\label{phi12firstquantdef}
\end{equation}
whose norm is
\begin{equation} \label{norm}
\mathcal{N}=1+\eta|\langle\phi_1|\phi_2\rangle|^2.
\end{equation}
According to the fact that the action of linear operators must comply with particle identity,  in a first quantization setting, one defines the action 
of a single particle operator $A$ on two-particle states as the operator $A^{(1)}$ which formally acts on the 2-particle state as
\begin{equation}
\label{LFscprod}
A^{(1)}\,\vert\phi_1,\phi_2\rangle=\vert A\phi_1,\phi_2\rangle\,+\,\vert\phi_1,A\phi_2\rangle\ .
\end{equation}
We shall refer to $A^{(1)}$ as the {\it 2-particle extended} (or just as the {\it extended}) {\it single-particle operator}.
By linearity and because of~\eqref{parity}, for one-particle projections $P_\psi:=\vert\psi\rangle\langle\psi\vert$ one finds
\begin{equation}
\label{LFproj1}
P^{(1)}_\psi\,\big\vert\phi_1,\phi_2\rangle=\vert \psi\,,\,\langle\psi\vert\phi_1\rangle\phi_2\,+\,\eta\,\langle\psi\vert\phi_2\rangle\phi_1\big\rangle\ ,
\end{equation}
whence the second vector suggests to define the following  \textit{reduction operator} $\Pi^{2\to1}_\psi$ from two particle states to one particle states, 
\begin{equation}
\label{LFproj2}
\Pi^{2\to1}_\psi\vert\phi_1,\phi_2\rangle=\langle\psi\vert\phi_1\rangle\,\vert\phi_2\rangle\,+\,\eta\,\langle\psi\vert\phi_2\rangle\,\vert\phi_1\rangle\ .
\end{equation}

Let us now consider an orthonormal basis $\{\vert\psi_k\rangle\}_{k}$ in the one-particle Hilbert space $\mathbb{H}$ and a subspace $\mathbb{K}$ linearly spanned by a subset $K$ of these vectors with associated orthogonal projector 
$\displaystyle P_{\mathbb{K}}=\sum_{k\in K}\vert\psi_k\rangle\langle\psi_k\vert$. Given a normalized two-particle vector state
\begin{equation}
\label{phi12}
\vert\Phi_{12}\rangle:=\frac{\vert\phi_1,\phi_2\rangle}{\sqrt{\mathcal{N}}}\ ,
\end{equation}
one can then associate with it a $\mathbb{K}$-reduced one-particle density matrix $\rho^{\mathbb{K}}_{\Phi}$ by
$$
\vert\Phi_{12}\rangle\langle\Phi_{12}\vert\mapsto
\sum_{k\in K}\Pi^{2\to1}_{\psi_k}\vert\Phi_{12}\rangle\langle\Phi_{12}\vert\big(\Pi^{2\to1}_{\psi_k}\big)^\dag
$$ 
and normalization of the right-hand-side. Using~\eqref{LFproj2} one gets
%VERSIONE ORIGINALE
%\begin{eqnarray}
%\label{LFredmat1}
%\hskip-1cm
%\rho^{\mathbb{K}}_\Phi&:=&
%\frac{\sum_{k\in K}\Pi^{2\to1}_{\psi_k}\vert\Phi_{12}\rangle\langle\Phi_{12}\vert\big(\Pi^{2\to1}_{\psi_k}\big)^\dag}{\sum_{k\in K}\|\Pi^{2\to1}_{\psi_k}\vert\Phi_{12}\rangle\|^2}\\
%\label{LFredmat2}
%\hskip-1cm
%&=&
%\frac{\langle\phi_2\vert P_{\mathbb{K}}\phi_2\rangle\,P_1\,+\,\langle\phi_1\vert P_{\mathbb{K}}\phi_1\rangle\,P_2+\eta\big(\langle\phi_1\vert P_{\mathbb{K}}\phi_2\rangle\,\vert\phi_1\rangle\langle\phi_2\vert+\langle\phi_2\vert P_{\mathbb{K}}\phi_1\rangle\,\vert\phi_2\rangle\langle\phi_1\vert\big)}{\langle\phi_2\vert P_{\mathbb{K}}\phi_2\rangle+\langle\phi_1\vert P_{\mathbb{K}}\phi_1\rangle+2\eta\mathcal{R}e\big(\langle\phi_1\vert P_{\mathbb{K}}\phi_2\rangle\langle\phi_2\vert\phi_1\rangle\big)}\ ,
%\end{eqnarray}
%where $P_{1,2}:=\vert\phi_{1,2}\rangle\langle\phi_{1,2}\vert$.
%VERSIONE ALTERNATIVA
\begin{eqnarray}
\label{LFredmat1}
%\hskip-1cm
\rho^{\mathbb{K}}_\Phi &:=&
\frac{1}{2\mathcal{N}_{\mathbb{K}}}\sum_{k\in K}\Pi^{2\to1}_{\psi_k}\vert\Phi_{12}\rangle\langle\Phi_{12}\vert\big(\Pi^{2\to1}_{\psi_k}\big)^\dag\\
\label{LFredmat2}
%\hskip-1cm
&=&
\frac{1}{2\mathcal{N}_{\mathbb{K}}}\Big(\langle\phi_2\vert P_{\mathbb{K}}|\phi_2\rangle\,\vert\phi_1\rangle\langle\phi_1\vert+\langle\phi_1\vert P_{\mathbb{K}}|\phi_1\rangle\,\vert\phi_2\rangle\langle\phi_2\vert \nonumber\\
&& \qquad +\eta\big(\langle\phi_1\vert P_{\mathbb{K}}|\phi_2\rangle\,\vert\phi_1\rangle\langle\phi_2\vert+\langle\phi_2\vert P_{\mathbb{K}}|\phi_1\rangle\,\vert\phi_2\rangle\langle\phi_1\vert\big)\Big)\ ,
\end{eqnarray}
with the normalization
\begin{equation}
\mathcal{N}_{\mathbb{K}}=\frac{1}{2}\sum_{k\in K}\|\Pi^{2\to1}_{\psi_k}\vert\Phi_{12}\rangle\|^2=\frac{1}{2}\langle\phi_1\vert P_{\mathbb{K}}|\phi_1\rangle+\frac{1}{2}\langle\phi_2\vert P_{\mathbb{K}}|\phi_2\rangle+\eta\mathcal{R}e\big(\langle\phi_1\vert P_{\mathbb{K}}|\phi_2\rangle\langle\phi_2\vert\phi_1\rangle\big) \ .
\end{equation}

Similarly to the case of bipartite states of distinguishable particles, the entanglement content of the bipartite state $\vert\Phi_{12}\rangle$ is then measured by the von Neumann entropy of the $\mathbb{K}$-reduced density matrix
\begin{equation}
\label{LFent}
E_{\mathbb{K}}(\Phi_{12})=S\big(\rho_\Phi^{\mathbb{K}}\big)=-{\rm Tr}\big(\rho_\Phi^{\mathbb{K}}
\log\rho^{\mathbb{K}}_\Phi\big)\ .
\end{equation}

\begin{remark}
\label{rem2}
Notice that the reduction illustrated above does not yield a density matrix supported by the subspace $\mathbb{K}$; rather, it is supported by the two-dimensional sub-space generated by $\vert\phi_{1,2}\rangle$ with matrix elements depending on $\mathbb{K}$.
As explicit in~\eqref{LFredmat1}, while the entanglement measured by~\eqref{LFent} does not depend on the chosen orthonormal basis of $\mathbb{K}$, it however depends on the sub-space $\mathbb{K}$.
\end{remark}
\begin{remark}
\label{rem2bis}
Because of the action of the reduction operator $\Pi^{2\to1}_\psi$, the $\mathbb{K}$-reduced density matrix \eqref{LFredmat1} does not reproduce the expectation values over $\vert\Phi_{12}\rangle$ of the extended operator $A^{(1)}$ starting from the single-particle operator $A$:
$$
\langle\Phi_{12}\vert A^{(1)}|\Phi_{12}\rangle\neq{\rm Tr}\big(\rho_\Phi^{\mathbb{K}}\,A\big) \; .
$$
Indeed, application of~\eqref{LFscprod} and~\eqref{LFredmat1} yields respectively
\begin{eqnarray}
\label{single-part-exp1}
\langle\Phi_{12}\vert A^{(1)}|\Phi_{12}\rangle&=& {1 \over \mathcal{N}} \Big[ \langle\phi_1\vert A|\phi_1\rangle+\langle\phi_2\vert A|\phi_2\rangle+2\eta\mathcal{R}e\big(\langle\phi_2\vert A|\phi_1\rangle\,\langle\phi_1\vert\phi_2\rangle\big) \Big]
%{1+\eta|\langle\phi_1\vert\phi_2\rangle|^2} 
\\
{\rm Tr}\big(\rho_\Phi^{\mathbb{K}}\,A\big)&=& {1 \over 2\mathcal{N}_{\mathbb{K}}} \Big[\langle\phi_1\vert A|\phi_1\rangle\langle\phi_2\vert\,P_{\mathbb{K}}|\phi_2\rangle+\langle\phi_1\vert\,P_{\mathbb{K}}\phi_1\rangle\langle\phi_2\vert A|\phi_2\rangle \; ,
\nonumber \\
&& \qquad  +2\eta\mathcal{R}e\big(\langle\phi_2\vert A|\phi_1\rangle\,\langle\phi_1\vert\,P_{\mathbb{K}}|\phi_2\rangle\big)\Big] \; ,
%\nonumber \\
%&& %\Big(\langle\phi_1\vert\,P_{\mathbb{K}}\phi_1\rangle+\langle\phi_2\vert\,P_{\mathbb{K}}\phi_2\rangle+2\eta|\langle\phi_1\vert\,P_{\mathbb{K}}\phi_2\rangle|^2\Big)^{-1} \: ,
\label{single-part-exp2}
\end{eqnarray}
which are in general different. 
Equivalence between expressions (\ref{single-part-exp1}) and (\ref{single-part-exp2}) can be established, for instance, if $\mathbb{K}=\mathbb{H}$, namely if $P_{\mathbb{K}}=\bf{1}$ and there is no sub-space projection but only single-particle reduction (note that in this case $\mathcal{N}_{\mathbb{H}}=\mathcal{N}$). There still remains, however, a factor $1/2$ of difference between \eqref{single-part-exp1} and \eqref{single-part-exp2}, which could be corrected by normalizing the symmetrized action~\eqref{LFscprod} of single-particle observables.
The physical origin of this factor lies in the fact that the trace in (\ref{single-part-exp2}) reproduces expectations of observables supported by the single particle operators
%on the untraced particle, 
$A\otimes 1$ or $1\otimes A$, and not by the symmetrized operator
%2-particle extended operator 
$A^{(1)}$. The latter is the only physically addressable single-particle operators in the setting of indistinguishable particles. From its definition in~\eqref{LFscprod}, it counts single particle contributions from both particles.
Moreover, in the $\mathbb{K}=\mathbb{H}$ case, the density matrix (\ref{LFredmat2}) of the no-label approach becomes
\begin{equation}
\label{LFredmat}
\rho_\Phi=\frac{|\phi_1\rangle\langle\phi_1|+|\phi_2\rangle\langle\phi_2|+\eta\big(\langle\phi_1\vert \phi_2\rangle\,\vert\phi_1\rangle\langle\phi_2\vert+\langle\phi_2\vert \phi_1\rangle\,\vert\phi_2\rangle\langle\phi_1\vert\big)}{2(1+\eta|\langle\phi_2\vert\phi_1\rangle|^2)},
\end{equation}
which is formally equal to the standard reduced density matrix obtainable by partial trace in the case of bipartite systems of distinguishable particles: that is, by considering $|\Phi_{12}\rangle$ defined as in (\ref{phi12firstquantdef}) and by tracing over the first or the second particle $\textnormal{Tr}_{(1)}|\Phi_{12}\rangle\langle\Phi_{12}|=\textnormal{Tr}_{(2)}|\Phi_{12}\rangle\langle\Phi_{12}|$.
%and, from a straightforward computation, equals the standard reduced density matrix obtainable by partial trace as in the case of bipartite systems of distinguishable particles, i.e. $\textnormal{Tr}_1|\Phi_{12}\rangle\langle\Phi_{12}|=\textnormal{Tr}_2|\Phi_{12}\rangle\langle\Phi_{12}|$. 
\end{remark}

\subsection{The two approaches compared}

We now consider the state $\vert\Phi_{12}\rangle=\vert\phi_1,\phi_2\rangle /\sqrt{\mathcal{N}}$ in~\eqref{phi12} with $\langle\phi_1\vert\phi_2\rangle=0$; in the second quantization
approach, $\vert\Phi_{12}\rangle=a^\dag_1a^\dag_2\vert 0\rangle$ where $a^\dag_a\vert 0\rangle=\vert\phi_a\rangle$, $a=1,2$. 
In this latter context, as already discussed in Section~\ref{sec3.1}, $\vert\Phi_{12}\rangle$ is separable with respect to the bipartition consisting of the two commuting sub-algebras $\mathcal{A}_1$, respectively $\mathcal{A}_2$ generated by $a_1,a^\dag_1$, respectively $a_2,a^\dag_2$. On the other hand, $\vert\Phi_{12}\rangle$  is of course entangled with respect to bipartitions obtainable by a Bogoljubov transformation that linearly combines the $a$-modes.

In the no-label approach, commuting extended single particle observables $O_1^{(1)}$ and $O_2^{(1)}$
come from observables on the single particle Hilbert space $\mathbb{H}$ such that $[O_1,O_2]=0$. Then, using~\eqref{LFscprod}, one finds that factorization as in~\eqref{1}, with $\tilde O_j$ substituted by $O_j^{(1)}$, occurs if and only if ($O_a=O_a^\dag$, $a=1,2$)
\begin{eqnarray}
\nonumber
\hskip-1cm
\langle\phi_1\vert O_1O_2|\phi_1\rangle\,+\,\langle\phi_2\vert O_1O_2|\phi_2\rangle\,+\,
2\eta\mathcal{R}e\Big(\langle\phi_1\vert O_1|\phi_2\rangle\langle\phi_2\vert O_2|\phi_1\rangle\Big)&=&
\langle\phi_1\vert O_1|\phi_1\rangle\langle\phi_1\vert O_2|\phi_1\rangle\\
\label{factorLF}
&+&
\langle\phi_2\vert O_2\phi_2\rangle\langle\phi_2\vert O_1\phi_2\rangle\ .
\end{eqnarray}
Given an orthonormal basis $\{\vert\phi_j\rangle\}_j$ in $\mathbb{H}$, consider the following cases:
\begin{enumerate}
\item
$O_1=\vert\phi_1\rangle\langle\phi_1\vert$, $O_2=\vert\phi_2\rangle\langle\phi_2\vert$: $O_1O_2=0$ and both sides of~\eqref{factorLF} vanish.\\ In this case, the two orthogonal projectors together with the identity  generate two commutative sub-algebras commuting between themselves. Then, the mean-value with respect to $\vert\Phi_{12}\rangle$  factorizes, as in the second quantization setting for the bipartition $(\mathcal{A}_1,\mathcal{A}_2)$ discussed before.
\item
$O_1=\vert\psi_+\rangle\langle\psi_+\vert$, $O_2=\vert\psi_-\rangle\langle\psi_-\vert$, with $\displaystyle\vert\psi_\pm\rangle=\frac{\vert\phi_1\rangle\pm\vert\phi_2\rangle}{\sqrt{2}}$: $O_1O_2=0$ and the left hand side of~\eqref{factorLF} equals $\displaystyle-\frac{\eta}{2}$, while the right hand side equals $\displaystyle\frac{1}{2}$.\\ 
In this case, the new projectors $O_{1,2}$ generate two commuting sub-algebras (together with the identity) over which the mean-values do not factorize for Bosons $(\eta=1$), while they do for Fermions ($\eta=-1$). This latter fact expresses the invariance of the singlet structure with respect to unitary rotations in the two-dimensional subspace to which it belongs.  
\item
$O_1=\vert\psi_+\rangle\langle\psi_+\vert$, $O_2=\vert\psi_-\rangle\langle\psi_-\vert$, with $\displaystyle\vert\psi_\pm\rangle=\frac{\vert\phi_1\rangle\pm\vert\phi_j\rangle}{\sqrt{2}}$, $j\geq 3$: $O_1O_2=0$ and the left hand side of~\eqref{factorLF} vanishes, while the right hand side equals $\displaystyle\frac{1}{4}$.\\
In this case, even the residual Fermionic separability of the previous point is lost.
\end{enumerate}

These examples show that, even within the no-label approach framework, separability/en\-tan\-glement of pure states are not absolute notions, since the factorization of mean values depends on the specific commuting observables one considers.

%These examples show that with respect to factorization of mean values, even in the no-label approach, separability of pure states is not an absolute notion, since factorizaton depends on the specific commuting observables one considers.

A further case considered in the no-label approach~\cite{LoFranco2016,LoFranco2017,Bellomo2017} is that of a single particle Hilbert space with a dicotomic space degree of freedom $L,R$ and an internal one $0,1$ as in the case of cold atoms confined within the left and right minima of a double-well potential and further distinguished by a pseudo-spin variable. Then, the single particle Hilbert space is linearly spanned by $4$ orthogonal vectors $\vert L,0\rangle$, $\vert L,1\rangle$, $\vert R,0\rangle$ and $\vert R,1\rangle$.
Within this framework, the entanglement of various bipartite states $\vert\Phi_{12}\rangle$ is considered by inspecting the reduction to left localized single-particle states obtained, acording to~\eqref{LFredmat1} and~\eqref{LFredmat2}, by restriction to the subspace $\mathbb{K}_L$ spanned by $\vert L,0\rangle$ and $\vert L, 1\rangle$:
\begin{equation}
\label{leftloc}
\rho_\Phi^L=\frac{\langle\phi_2\vert P_L|\phi_2\rangle\vert\phi_1\rangle\langle\phi_1\vert+\langle\phi_1\vert P_L|\phi_1\rangle\vert\phi_2\rangle\langle\phi_2\vert+\eta\big(\langle\phi_1\vert P_L|\phi_2\rangle\vert\phi_1\rangle\langle\phi_2\vert+\langle\phi_2\vert P_L|\phi_1\rangle\vert\phi_2\rangle\langle\phi_1\vert\big)}{\langle\phi_2\vert P_L|\phi_2\rangle+\langle\phi_1\vert P_L|\phi_1\rangle+2\eta\mathcal{R}e\big(\langle\phi_1\vert P_L|\phi_2\rangle\langle\phi_2\vert\phi_1\rangle\big)}\ ,
\end{equation}
where we have defined $P_L:=P_{\mathbb{K}_L}=\vert L,0\rangle\langle L,0\vert\,+\,\vert L,1\rangle\langle L,1\vert$ to simplify the notation.

As observed before, $\rho^L_\Phi$ is not the partial trace over the right degrees of freedom that reduces $\vert\Phi_{12}\rangle\langle\Phi_{12}\vert$ 
to a density matrix for the left degrees of freedom, and is thus crucial to understand what the von Neumann entropy of $\rho^L_{\Phi_{12}}$ can tell about the entanglement of $\vert\Phi_{12}\rangle$.
As in~\cite{LoFranco2016,LoFranco2017,Bellomo2017}, we shall consider the following cases.
\begin{enumerate}
\item
$\vert\Phi_{12}\rangle=\vert L,0;R,1\rangle$, with single particle vector states $\vert\phi_1\rangle=\vert L,0 \rangle$, $\vert\phi_2\rangle=\vert R,1 \rangle$: then,
\begin{equation}
\rho_\Phi^L=\vert R,1\rangle\langle R,1\vert\ \Longrightarrow\ E_L(\Phi_{12})=0
\label{case1EL}
\end{equation}
so that $\vert\Phi_{12}\rangle$ is deemed separable in the no-label approach.
\item
$\displaystyle\vert\Phi_{12}\rangle=\frac{1}{\sqrt{2}}\vert L,0;L,0\rangle$; then,
\begin{equation}
\rho_\Phi^L=\vert L,0\rangle\langle L,0\vert\ \Longrightarrow\ E_L(\Phi_{12})=0\ .
\label{case2EL}
\end{equation}
Then, again $\vert\Phi_{12}\rangle$ seems separable in the no-label approach.
\item
$\vert\Phi_{12}\rangle=\vert L,0;L,1\rangle$; then,
\begin{equation}
\rho_\Phi^L=\frac{1}{2}\Big(\vert L,0\rangle\langle L,0\vert\,+\,\vert L,1\rangle\langle L,1\vert\Big)
\ \Longrightarrow\ E_L(\Phi_{12})=1\ ,
\label{case3EL}
\end{equation} 
so that $\vert\Phi_{12}\rangle$ is established as maximally entangled by the no-label approach.
\end{enumerate}

We now compare the separability and entanglement obtained above within the no-label approach with the factorization of mean-values.

Let us then consider the orthogonal single particle projectors $P_1,P_2$;
using~\eqref{LFscprod0} and~\eqref{LFscprod} one obtains
\begin{eqnarray*}
\langle\Phi_{12}\vert (P_1)^{(1)}(P_2)^{(1)}|\Phi_{12}\rangle&=&\frac{1}{\mathcal{N}}\Big(
\langle\phi_1\vert P_1|\phi_1\rangle\langle\phi_2\vert P_2|\phi_2\rangle\,+\,
\langle\phi_1\vert P_2|\phi_1\rangle\langle\phi_2\vert P_1|\phi_2\rangle\\
&+&\eta\Big(\langle\phi_2\vert P_1|\phi_1\rangle\langle\phi_1\vert P_2|\phi_2\rangle\,+\,
\langle\phi_2\vert P_2|\phi_1\rangle\langle\phi_1\vert P_1|\phi_2\rangle
\Big)\Big)\\
\langle\Phi_{12}\vert (P_{1,2})^{(1)}|\Phi_{12}\rangle&=&\frac{1}{{\mathcal{N}}}\Big(
\langle\phi_1\vert P_{1,2}|\phi_1\rangle\,+\,
\langle\phi_2\vert P_{1,2}|\phi_2\rangle\\
&+&\eta\Big(\langle\phi_1\vert P_{1,2}|\phi_2\rangle\langle\phi_2\vert\phi_1\rangle\,+\,
\langle\phi_2\vert P_{1,2}|\phi_1\rangle\langle\phi_1\vert\phi_2\rangle
\Big)\Big)\ ,
\end{eqnarray*}
where $\mathcal{N}$ is the normalization factor in equation \eqref{norm}. 
In the three examples considered above one thus gets
\begin{enumerate}
\item
$\vert\Phi_{12}\rangle=\vert L,0;R,1\rangle$, $\mathcal{N}=1$: let
$P_{1,2}=P^L_\pm:=\vert L,\pm\rangle\langle L,\pm\vert$ with $\displaystyle\vert L,\pm\rangle:=\frac{\vert L,0\rangle\pm\vert L,1\rangle}{\sqrt{2}}$; then,
$$
\langle\Phi_{12}\vert (P^L_+)^{(1)}(P^L_-)^{(1)}|\Phi_{12}\rangle=0\ ,\quad
\langle\Phi_{12}\vert (P^L_+)^{(1)}|\Phi_{12}\rangle\langle\Phi_{12}\vert (P^L_-)^{(1)}|\Phi_{12}\rangle=\frac{1}{4}\ .
$$
The state is thus entangled with respect to the single particle commuting 
observables $P_\pm^L$, while \eqref{case1EL} gives a vanishing entanglement entropy.
\item
$\vert\Phi_{12}\rangle=\vert L,1;L,1\rangle$, $\mathcal{N}=2$: let
$P_{1,2}=P^L_\pm:=\vert L,\pm\rangle\langle L,\pm\vert$ as before; then,
$$
\langle\Phi_{12}\vert (P^L_+)^{(1)}(P^L_-)^{(1)}|\Phi_{12}\rangle=\frac{1}{2}\ ,\quad
\langle\Phi_{12}\vert (P^L_+)^{(1)}|\Phi_{12}\rangle\langle\Phi_{12}\vert (P^L_-)^{(1)}|\Phi_{12}\rangle=1\ .
$$
The state is thus entangled with respect to the single particle commuting 
observables $P_\pm^L$, while \eqref{case2EL} indicates separability.
\item
$\vert\Phi_{12}\rangle=\vert L,0;L,1\rangle$, $\mathcal{N}=1$: let $P^L_a=\vert L,a\rangle\langle L,a\vert$, $a=0,1$; then,
$$
\langle\Phi_{12}\vert (P^L_0)^{(1)}(P^L_1)^{(1)}|\Phi_{12}\rangle=1\ ,\quad
\langle\Phi_{12}\vert (P^L_0)^{(1)}|\Phi_{12}\rangle\langle\Phi_{12}\vert (P^L_1)^{(1)}|\Phi_{12}\rangle=1\ .
$$
The state is thus separable with respect to the commuting sub-algebras $\mathcal{A}_{0,1}$ generated by the projections $P^L_{0,1}$ together with the single-particle identity operator, while \eqref{case3EL} characterizes this state as maximally entangled, regardless of the context.
\end{enumerate}

\section{Conclusions}

We argued that a universal and physically natural notion of separability, and thus of entanglement,  is based on the factorization of mean values of commuting observables with respect to pure states.\\
This characterization is nothing but a mathematical expression of ``absence of classical correlations'' and covers the standard setting of distinguishable particles, while being easily extensible to identical particle systems of Fermions and Bosons. An immediate consequence is that separability and entanglement become properties not of the states alone, but of  
of the states with respect to specified commuting sub-algebras of observables.\\
We tested this notion against a recent novel point of view regarding the description of identical particles, that we referred to as no-label approach, based not upon the symmetrization or antisymmetrization of vector states, rather on the introduction of a suitably constructed single particle density matrix whose von Neumann entropy is taken as an absolute measure of entanglement of two-particle states.\\
We showed that states with non-zero reduced von Neumann entropy turn out to be separable (that is factorizing) over certain commuting sub-algebras of observables and, vice versa, states with vanishing reduced von Neumann entropy are instead entangled, that is non-factorizing, in relation to other commuting sub-algebras. These examples indicate that, even within the no-label formalism, one cannot assign an absolute figure of merit to quantify entanglement, without specifying the type of observables one allows.
\bigskip

\noindent
Acknowledgments:\quad FF and UM acknowledge support from the H2020 CSA Twinning project No. 692194, ``RBI-T-WINNING'' and from the Croatian Science Fund Project No. IP-2016-6-3347.

\end{document}